\documentclass[prl,twocolumn,superscriptaddress,preprintnumbers,a4paper,amsmath,amssymb,showpacs,floatfix]{revtex4}
\usepackage{graphicx}
\usepackage{bm}

\begin{document}

\newcommand{\inA}{\mbox{\AA$^{-1}$}}
\newcommand{\q}{$\delta$}
\newcommand{\qlMn}{$q_{l}^{\rm Mn}$}
\newcommand{\qlDy}{$q_{l}^{\rm Dy}$}
\newcommand{\qlTb}{$q_{l}^{\rm Tb}$}
\newcommand{\qmMn}{$q_{m}^{\rm Mn}$}
\newcommand{\qmDy}{$q_{m}^{\rm Dy}$}
\newcommand{\qmTb}{$q_{m}^{\rm Tb}$}
\newcommand{\qMn}{$q^{\rm Mn}$}
\newcommand{\qmn}{$\delta_{m}^{\rm Mn}$}
\newcommand{\qdy}{$q^{Dy}$}
\newcommand{\dmn}{$\delta^{\rm Mn}$}
\newcommand{\ddy}{$\delta^{\rm Dy}$}
\newcommand{\bb}{$\mathbf{b}^{*}$}
\newcommand{\qc}{$\delta_{m}=\frac{1}{4}$}
\newcommand{\qcs}{$\delta_{l}=\frac{1}{2}$}
\newcommand{\TNMn}{$T_{N}$}
\newcommand{\TNDy}{$T_{N}^{\rm Dy}$}
\newcommand{\TlDy}{$T_{l}^{\rm Dy}$}
\newcommand{\TNTb}{$T_{N}^{\rm Tb}$}
\newcommand{\Tn}{$T_{N}$}
\newcommand{\TN}{$T_{N}$}
\newcommand{\Tc}{$T_{c}$}
\newcommand{\etal}{\textit{et al.}}
\newcommand{\degg}{$^{\circ}$}
\newcommand{\pa}{$\mathbf{P}\|a$}
\newcommand{\pb}{$\mathbf{P}\|b$}
\newcommand{\pc}{$\mathbf{P}\|c$}
\newcommand{\hh}{$\mathbf{H}$}
\newcommand{\Ps}{$P_{s}$}
\newcommand{\Psvect}{$\mathbf{P}_{s}$}
\newcommand{\mc}{$\mu C/m^{2}$}

\title{Enhanced ferroelectric polarization by induced Dy spin-order in multiferroic DyMnO$_{3}$}

\author{O.~Prokhnenko}
\affiliation{Hahn-Meitner-Institut, Glienicker Str.~100, D-14109 Berlin, Germany}

\author{R.~Feyerherm}
\affiliation{Hahn-Meitner-Institut, c/o BESSY, D-12489 Berlin, Germany}

\author{E.~Dudzik}
\affiliation{Hahn-Meitner-Institut, c/o BESSY, D-12489 Berlin, Germany}

\author{S.~Landsgesell}
\affiliation{Hahn-Meitner-Institut, Glienicker Str.~100, D-14109 Berlin, Germany}

\author{N.~Aliouane}
\affiliation{Hahn-Meitner-Institut, Glienicker Str.~100, D-14109 Berlin, Germany}

\author{L.C.~Chapon}
\affiliation{ISIS, Rutherford Appleton Laboratory, Chilton, Didcot, OX11, 0QX United Kingdom}

\author{D.~N.~Argyriou}
\affiliation{Hahn-Meitner-Institut, Glienicker Str.~100, D-14109 Berlin, Germany}

\date{\today}
\pacs{61.12.Ld, 61.10.-i, 75.30.Kz, 75.47.Lx, 75.80.+q}

\begin{abstract}
Neutron powder diffraction and single crystal x-ray resonant magnetic scattering measurements suggest that Dy plays an active role in enhancing the
ferroelectric polarization in multiferroic DyMnO$_{3}$ above \TNDy=~6.5~K. We observe the evolution of an incommensurate ordering of Dy moments with the same periodicity as the Mn spiral ordering. It closely tracks the evolution of the ferroelectric polarization which reaches a maximum value of 0.2~\mc. Below \TNDy, where Dy spins order commensurately, the polarization decreases to values similar for those of TbMnO$_{3}$.
\end{abstract}

\maketitle 

The strong coupling between ferroelectricity and magnetism in modern multiferroics has offered a new paradigm of magnetoelectric materials. It has stimulated a search for new multiferroics and highlighted the need for a deeper understanding of the physics behind these exciting materials. Although the requirements of magnetism and ferroelectricity are chemically incompatible\cite{Hill}, in these new multiferroics spin frustration\cite{kimura2} leads to complex magnetic arrangements that can break inversion symmetry\cite{kimura,lotter}. The strong coupling provided by frustration in the perovskite manganites $R$MnO$_{3}$ leads to a number of magneto-electric phenomena. For example, compounds with $R$=Tb and Dy exhibit flops of the direction of the spontaneous polarization (\Psvect) with applied field (\hh) while ferroelectricity is observed only under a magnetic field in $R$=Gd\cite{kimura,goto,kimura3,fiebig:review}. Of these perovskites, $R$=Dy exhibits the largest value of \Ps=0.2~$\mu C/m^{2}$ ($\sim$ three times larger than that for $R=$Tb) and a giant magnetocapacitance effect\cite{goto}.

In these materials, a phenomenological treatment of the coupling of a uniform electric polarization $\mathbf{P}$ to an inhomogeneous magnetization $\mathbf{M}$ leads to a term linear in the gradient $\nabla\mathbf{M}$, the so-called Lifshitz invariant, that is allowed only in systems with broken inversion symmetry\cite{mostovoy:067601}. This model is consistent with neutron diffraction experiments on TbMnO$_3$ that showed that a spiral arrangement of Mn-spins within the $bc$-plane develops at $T_l$ = 28~K, coinciding with the onset of \Psvect\cite{kenz}. Although in this picture the contribution to \Psvect~from the magnetic ordering of $R$-ions is ignored, their role is underscored as a source of magnetic anisotropy that is required to predict the correct direction of \Psvect~under an applied magnetic field\cite{mostovoy:067601}.

In this Letter we present measurements on DyMnO$_{3}$ suggesting that Dy plays an active role in enhancing the ferroelectric polarization in multiferroic DyMnO$_{3}$ above \TNDy=~6.5~K. We have combined neutron powder diffraction and single crystal x-ray resonant magnetic scattering to investigate the evolution of magnetism in DyMnO$_{3}$ with temperature. We find that the Dy moments order in an incommensurate (ICM) structure with the same periodicity as the Mn moments below a temperature \TlDy= 15~K. The transition from the commensurate (CM) Dy ordering below \TNDy~to the ICM state is associated with an enhancement of \Ps~just above \TNDy~to a value approximately twice that found for the maximum \Ps~in TbMnO$_{3}$. The CM-ICM transition shows a large hysteresis in which \Ps~and the intensity of magnetic reflections arising from the ICM Dy magnetic ordering exhibit a similar behavior. Our work suggests a magneto-strictive coupling between Mn and Dy spins giving rise to an enhanced \Ps~above \TNDy~compared to other perovskite multiferroics. This mechanism would add to the ferroelectric polarization that arises from the Mn spin-spiral.

In the multiferroic perovskite manganites it is found that for $R$=Tb below \TN$\sim$41~K, Mn-spins order first along the $b-$axis in an ICM sinusoidal arrangement with propagation vector $\mathbf{\tau}^{Mn}$ = (0 0.28...0.29 0)\cite{kimura2,Quezel}. A similar behavior with $\mathbf{\tau}^{Mn}$ = (0 0.36...0.385 0) is expected from x-ray measurements for DyMnO$_3$\cite{kimura2}. Second harmonic lattice reflections ($q=2\tau$) associated with the magnetic ordering have been observed to arise from a coupling of the ICM magnetic ordering to the lattice via a quadratic magneto-elastic coupling\cite{PW,walker}. Below $T_l$=28~K for TbMnO$_3$, an additional component of the Mn magnetic moment along the $c$-axis, in phase quadrature with the component along $b$, gives rise to a spiral magnetic ordering and breaks inversion symmetry, leading to the observation of \Psvect\space along the $c-$axis. In this regime Mn spins induce an ICM ordering of Tb-spins with the same propagation vector\cite{kenz}. A similar transition occurs for $R$=Dy at $T_l$=18K\cite{goto} as illustrated in Fig.~1(a). Below $T_N^{R}<$10~K, Tb and Dy magnetic moments order separately with propagation vectors $\tau^{Tb}$ = (0 0.42 0) and  $\tau^{Dy}$=(0 $\frac{1}{2}$ 0)\cite{kajimoto,feyerherm}.  

\begin{figure}[bt!]
\begin{center}
\includegraphics[scale=1.0]{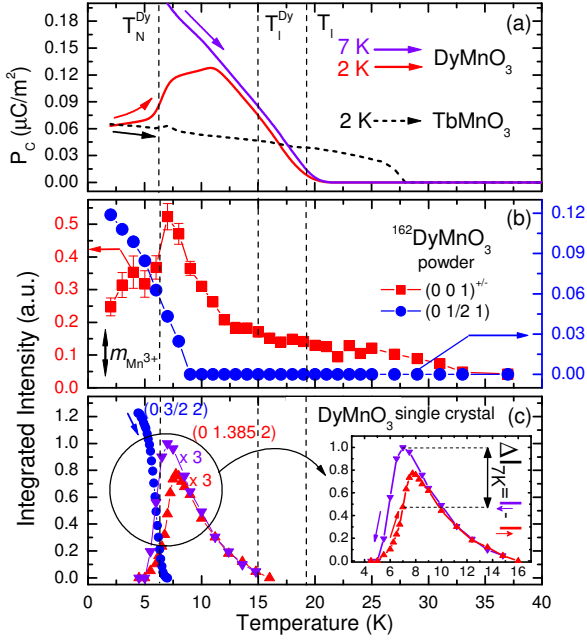}
\caption{(Color online) Temperature dependences of (a) electric polarization along the $c$-axis for DyMnO$_3$ and TbMnO$_3$ (from Goto $\it{et~al}$.\cite{goto}), (b) integrated (neutron) intensities of magnetic A-type and (0 $\frac{1}{2}$ 1) reflections measured on heating ($\updownarrow m_{Mn^{3+}}$ shows the intensity corresponding to a 4~$\mu_B$ Mn$^{3+}$ magnetic moment for this reflection; shaded area corresponds to the magnetic intensity on top of the Mn$^{3+}$ saturated  moment), (c) temperature dependent integrated (synchrotron) intensities of selected Dy commensurate (heating) and incommensurate (heating and cooling) superlattice reflections. The region around \TNDy is shown in the insert. Here, $\Delta$I highlights the difference between intensities on cooling and heating at 7~K.}
\label{fig1}
\end{center}
\end{figure}

To investigate the evolution of the magnetic ordering of DyMnO$_{3}$ with neutron powder diffraction (NPD) we used a 0.65g polycrystalline $^{162}$DyMnO$_3$ sample prepared from a mixture of isotope enriched $^{162}$Dy$_2$O$_3$ oxide (94.4\% enrichment) and Mn$_2$O$_3$, using standard solid state synthesis methods. Isotopic $^{162}$Dy was chosen for its smaller neutron absorption cross section ($\sigma_{a}$) compared to that of natural Dy\cite{absorption}. NPD data were measured from this sample between 2-300~K on the GEM diffractometer at the ISIS-facility, Rutherford-Appleton Laboratory. The data were analyzed with the FullProf refinement package\cite{carvajal}. Synchrotron x-ray diffraction measurements on a DyMnO$_{3}$ single crystal were conducted at the 7~T multipole wiggler beamline MAGS, operated by the Hahn-Meitner-Institut at the synchrotron source BESSY in Berlin. Details on the crystal growth, beamline and experimental procedure have been reported previously\cite{feyerherm,dudzik}. Low field (500 Oe) magnetization of the polycrystalline and single crystal samples was measured in a SQUID magnetometer between 4 - 100~K. The data showed \TNDy\space = 9 and 6~K for the polycrystalline and single crystal sample, respectively. The value of \TNDy\space of our single crystal is in good agreement with published data\cite{kimura2} and may suggest that the higher value obtained for the powder sample may arise from a small non-stoichiometry. This assumption is corroborated by a difference in the lattice constants between our powder sample and the single crystal at 300~K\cite{feyerherm}.

\begin{figure}[bt!]
\begin{center}
\includegraphics[scale=1.0]{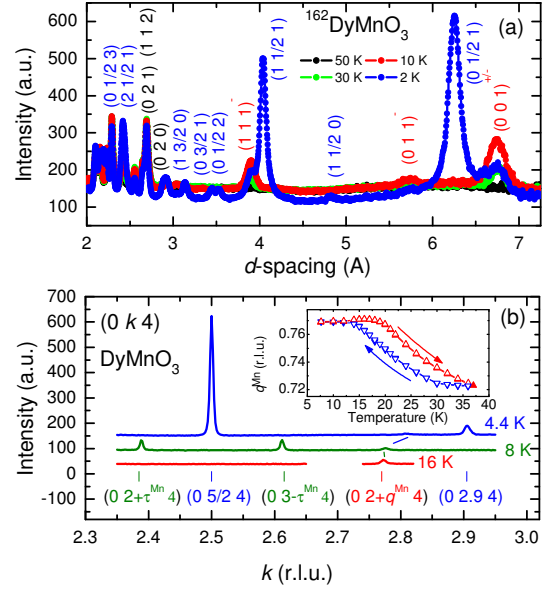}
\caption{(Color online) (a) Neutron diffraction patterns obtained from the 24$^\circ$-45$^\circ$ bank of the GEM diffractometer at 2, 10, 30 and 50~K. (b) $k$-scans along $(0~k~4)$ at various temperatures measured with a synchrotron x-ray energy close to the Dy-L$_3$ absorption edge. No polarization analysis was employed here in order to allow for simultaneous observation of structural and magnetic superstructure reflections. Inset: temperature dependence of the lattice modulation wave vector obtained from a superstructure reflection of type $(0~k~3)$.}
\label{fig2}
\end{center}
\end{figure}

Turning first to the NPD measurements we found that $^{162}$DyMnO$_3$ crystallized with the orthorhombically distorted perovskite structure (space group $Pbnm$). Cooling the sample below $T_N \sim $ 40~K we find a series of magnetic satellites arising from Mn spin-ordering (see Fig.~\ref{fig2}(a)) similar to the observations of Quezel \etal~\cite{Quezel} on TbMnO$_3$. Using their notation, these $(hkl) \pm \tau$ reflections are characterized as A-type where $h+k = even$ and $l = odd$ and $\tau=(0~\tau^{Mn}_y~0)$ is an incommensurate propagation vector along the $b^*$-axis. An increase of satellite intensities (and $\tau^{Mn}_y$) with decreasing temperature is accompanied by the appearance of G-type satellites ($h+k = odd$ and $l = odd$) below $\sim$15~K, where a non zero value of \Ps\space is observed as shown in Fig.~\ref{fig1}(a)\cite{goto,kimura3}. Below \TNDy\space the NPD data show an intensity reduction of the A-type satellites, which coincides with the appearance of a CM Dy magnetic order with propagation vector $\tau^{Dy}$ = (0 $\frac{1}{2}$ 0) (see Figs.~\ref{fig1}b, ~\ref{fig2}a)\cite{feyerherm}. This transition coincides with a sharp decrease of \Ps. The G-type reflections are too weak for a quantitative comparison. Rietveld analysis of the NPD data between \TNDy$<T<$13~K on the assumption of pure Mn spin-ordering (see below) leads to unphysically large moments for Mn$^{3+}$ ($> 4\mu_{B}$/Mn) (Fig.~\ref{fig1}(b)). The rapid increase of the magnetic intensity below 13~K (shaded area in Fig.~\ref{fig1}(b)) and its decrease below \TNDy\space suggests an additional  magnetic contribution to the intensity of these reflections from Dy. As can be seen in Fig.~\ref{fig2}(a), the magnetic reflections are relatively broad and below (Fig.~3, caption) we shall argue that this probably arises from a distribution of the propagation vectors.

In order to probe directly the magnetic contribution of Dy to the intensities of these ICM  magnetic reflections, we have used single crystal resonant x-ray scattering at an x-ray energy of 7.794~keV, slightly above the Dy-L$_3$ absorption edge. With this resonant condition, a survey of reciprocal space was carried out at different temperatures.  Fig.~\ref{fig2}(b) shows scans along the $(0~k~4)$ direction as examples. At 16~K only the Mn induced structural modulation $q^{Mn}$ is observed, producing the $(0~2.77~4)$ Bragg reflection. At 8~K a pair of relatively strong additional $(0~2.385~4)$ and $(0~2.615~4)$ reflections appears, consistent with the magneto-elastic coupling $q^{Mn}=2\tau^{Mn}$. At 4.4~K the latter reflections have disappeared, in contrast to the associated peak from the Mn induced structural modulation that decreases in intensity and is shifted towards a $q^{Mn}_y$ value of 0.81. The extinct reflections are replaced by significantly stronger Bragg reflections at $(0~2.5~4)$ and $(0~2.905~4)$ which correspond to the CM magnetic ordering of the Dy moments and the associated incommensurate lattice modulation $q^{Dy}_y = 0.905 \not = 2 \tau^{Dy}_y$ respectively. This transition has been discussed recently elsewhere\cite{feyerherm}.

\begin{figure}
\begin{center}
\includegraphics[scale=0.8]{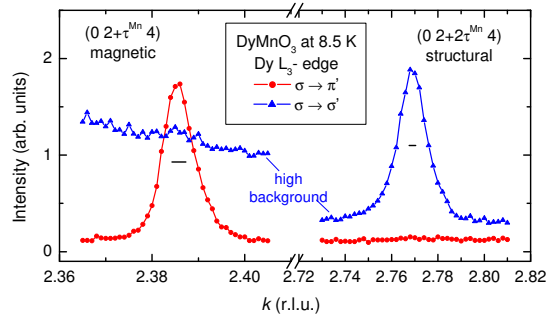}
\caption{(Color online) Linear polarization analysis of (left) the $(0~2.385~4)$ and (right) the $(0~2.77~4)$ superstructure reflections at 8.5~K. The intensity variation as a function of the polarization analyzer configuration shows that $(0~2.385~4)$ is of magnetic and $(0~2.77~4)$ of structural origin. The line widths of both reflections are significantly enhanced with respect to the instrumental resolution that is marked by horizontal bars. Since the line width of the second order reflection $(0~2.77~4)$ is roughly twice as large as that of the first order $(0~2.385~4)$, the broadening is more likely due to an inhomogeneous distribution of $\tau$values in the sample rather than to reduced coherence lengths.}
\label{fig3}
\end{center}
\end{figure}

In order to identify the nature of the Bragg reflections associated with $\tau^{Mn}$ above \TNDy~we employed linear polarization analysis\cite{feyerherm}. Figure~\ref{fig3} shows the dependence of the intensities of both the $(0~2.385~4)$ and the related $(0~2.77~4)$ reflections on the polarization analyzer configuration ($\sigma\rightarrow\sigma^{\prime}$ vs. $\sigma\rightarrow\pi^{\prime}$) at 8.5~K. The characteristic behavior shows that the former is of magnetic and the latter of structural origin. Tuning the x-ray energy to a value 20 eV below the Dy-L$_3$ absorption edge leads to a reduction of the intensity of the magnetic $(0~2.385~4)$ reflection measured in $\sigma\rightarrow\pi^{\prime}$ configuration by a factor of 40. This resonance enhancement shows that the $(0~2.385~4)$ reflection is due to ordered Dy magnetic moments and that any contribution of the ordered Mn moments to this reflection is negligible.
 
Figure \ref{fig1}(c) shows the temperature dependence of the integrated intensities of two particularly strong Bragg reflections, $(0~1.385~2)$ and $(0~1.5~2)$, related to $\tau^{Mn}$ and $\tau^{Dy}$, respectively. The half-integer reflection, associated with the CM magnetic ordering of the Dy moments, vanishes above \TNDy\space = 6.5 K. Simultaneously, the intensity of the ICM reflection increases steeply on heating. At the same temperature \Psvect$\|c$ increases to approximately twice its value compared to that at temperatures just below \TNDy\space (Fig.~\ref{fig1}(a)). On further heating, the $(0~1.385~2)$ Bragg intensity passes a maximum around 8~K, decreases monotonically with a concave curvature - typical for an induced magnetic moment, and finally vanishes above $T_{l}^{Dy} = 15$~K. In addition, it exhibits a significant hysteresis (see inset in Fig.~\ref{fig1}(c)). On cooling from 16~K, the $(0~1.385~2)$ Bragg intensity measured at 7~K is about twice as large as the intensity obtained on heating from 4.5 to 7~K. Monitoring the count rate at fixed $k = 1.385$ while sweeping the temperature at various rates (0.05...1~K/min.) we verified that whenever one does not cross \TNDy, the cooling and heating curves above \TNDy~are reversable and the hysteresis is independent of the heating/cooling rate. This behavior of Dy induced ordering strongly resembles the hysteresis of the electric polarization shown in Fig.~\ref{fig1}(a), where a factor of 2 difference is shown in the cooling and heating curves at 7K, the same difference that we find in the hysteresis of the intensity of the $(0~1.385~2)$ reflection. Finally, the insert in Fig.~\ref{fig2}(b) shows the temperature dependence of $q^{Mn}$ measured in two successive heating and cooling cycles between 7.5 and 40~K. Clearly, a significant hysteresis of about 5~K in width is observed in the temperature dependence. Neither on cooling nor on heating any sharp lock-in transition is observed. The edge of the heating curve, however, is consistent with the reported value for $T_l$ = 18~K.

Having directly established the behavior of Dy-spins above and below \TNDy~we turn our attention to models of magnetic ordering for Dy that can be obtained from the analysis of the NPD data. In the CM regime below \TNDy~the diffraction data indicates a cell doubling along $b$ arising from antiferromagnetic Dy spin-ordering. Rietveld analysis shows that the Dy moments with absolute value $m^{Dy}=6.2(6)\mu_B$/Dy lie in the $ab$ plane, being canted $30(\pm 10)^\circ$ from the $b$ axis, and are stacked antiferromagnetically along the c-axis.

In the analysis of the induced magnetic ordering of Dy-spins at 10~K, above \TNDy, we have taken solutions of the magnetic structure that are allowed by the $\Gamma_2 \otimes \Gamma_3$ representation of the $Pbnm$ space group indicated by the spiral arrangements of Mn-spins in TbMnO$_{3}$\cite{kenz,brinks}. In the case where two irreproducible representations are coupled\cite{kenz}, the $R$ moments are allowed to have components along the three principal crystallographic directions denoted as $\mathbf{m}^{Dy}=$(a$_x$, f$_y$, f$_z$)\cite{brinks}. Our Rietveld analysis shows that the magnetic moment is  $\mathbf{m}^{Dy}=$(0.4(6),2.5(2),0.0(3))~$\mu_B$/Dy. The maximum value of the moment here is approximately twice as much as that found for the induced moment of Tb in TbMnO$_{3}$ ($\mathbf{m}^{Tb}=$(1.2,0,0)$\mu_B$/Tb\cite{kenz}). In these refinements we assumed the same magnetic symmetry for Mn as that reported for TbMnO$_{3}$\cite{kenz}. 

The CM structure of Dy-spins below \TNDy~preserves the mirror plane perpendicular to the $c$ axis and thus obviously does not allow for a contribution to \Psvect~along the $c-$axis. Polarization measurements show that this spin ordering also does not contribute to \Psvect~along the other principal crystallographic directions (\pa$\sim$\pb$\sim0$)\cite{kimura3}. For the region just above \TNDy~our data is consistent with a sinusoidal modulation of the Dy-spins along the $b-$axis. The absence of a $c-$axis component would exclude the presence of a Dy $bc$-spiral as found for the case of Mn spin-ordering in TbMnO$_{3}$. Thus, on the basis of symmetry the induced Dy-spin order alone is unlikely to result in a ferroelectric polarization along the $c-$axis above \TNDy.   

Compared to $R$=Tb or (Eu,Y), the behavior of DyMnO$_{3}$ is unusual, as \Ps~just above \TNDy~ is three times larger for comparable conditions and shows a large hysteresis. Indeed Goto \etal\cite{goto} find almost a 50\% difference in \Ps~at 7~K obtained after cooling to 2 and 7~K, a behavior that is similar to the hysteresis of the magnetic intensity from the induced Dy spin-order (Fig.~\ref{fig1}(c)). Below \TNDy, \Ps~decreases to a value of 0.06\mc~ which is similar to values  found for non-magnetic $R$-ions and $R$=Tb at 2K. This would indicate that in this regime ferroelectricity arises from a similar coupling in all of these materials. 

Our measurements indicate that there is an additional (or alternate) mechanism by which Dy-spins can provide a significant enhancement of \Ps~above the value expected from a Mn spiral alone. Although our measurements can not provide a detailed microscopic model of the lattice distortions that arise from the magnetic ordering of Mn and Dy, we can exclude a mechanism based on a $bc$ Dy spin-spiral. The temperature hysteresis of \Ps~and the induced Dy magnetic reflection occur over the same region, suggesting a close coupling between these two quantities. The lattice distortion that arises from the Dy induced ordering is significantly larger in magnitude and different in nature than those for $R$=Tb. The hysteretic behavior of \Ps~and the intensity of the ICM Dy magnetic reflections suggests that the enhancement of \Ps~above \TNDy~arises from a magneto-strictive coupling between Mn and Dy spins. Indeed this may be a peculiarity of the single ion anisotropy of Dy since it is a Kramers ions and Tb in contrast is not. Clearly the magneto-strictive lattice distortions must be conform to $\Gamma_2 \otimes \Gamma_3$ symmetry.

The present results are important for a full description of the magneto-electric coupling in perovskite manganites. While $R$-spins thus far have been perceived to be spectators to ferroelectricity, here we show that in DyMnO$_{3}$ the induced magnetic ordering of Dy coincides with a region of enhanced polarization compared to other $R$MnO$_{3}$ manganites, while below \TNDy~when Dy spins order commensurately \Ps~is sharply reduced. We argue that these observation suggest a peculiar coupling of the lattice that enhances \Ps~when Dy-spins are ordered incommensurately above \TNDy.

The authors have benefited from discussions with M. Mostovoy, B. Harris and T. Kimura. Construction of the beamline MAGS has been founded by the BMBF via the HGF-Vernetzungsfonds under contracts No.~01SF0005 and 01SF0006. S. Landsgesell thanks the DFG for financial support via contract No.~AR 613/1-1.


\begin{thebibliography}{19}
\expandafter\ifx\csname natexlab\endcsname\relax\def\natexlab#1{#1}\fi
\expandafter\ifx\csname bibnamefont\endcsname\relax
  \def\bibnamefont#1{#1}\fi
\expandafter\ifx\csname bibfnamefont\endcsname\relax
  \def\bibfnamefont#1{#1}\fi
\expandafter\ifx\csname citenamefont\endcsname\relax
  \def\citenamefont#1{#1}\fi
\expandafter\ifx\csname url\endcsname\relax
  \def\url#1{\texttt{#1}}\fi
\expandafter\ifx\csname urlprefix\endcsname\relax\def\urlprefix{URL }
\fi
\providecommand{\bibinfo}[2]{#2}
\providecommand{\eprint}[2][]{\url{#2}}

\bibitem[{\citenamefont{Hill}(2000{\natexlab{a}})\citenamefont{Hill}}]{Hill}
\bibinfo{author}{\bibfnamefont{N.A.}~\bibnamefont{Hill},
  \bibinfo{journal}{J.Phys. Chem. B}
   \textbf{\bibinfo{volume}{104}}, \bibinfo{pages}{6694}
  (\bibinfo{year}{2000}{\natexlab{a}}}).

\bibitem[{\citenamefont{Kimura et~al.}(2003{\natexlab{b}})\citenamefont{Kimura,
  Ishihara, Shintani, Arima, Takahashi, Ishizaka, and Tokura}}]{kimura2}
\bibinfo{author}{\bibfnamefont{T.}~\bibnamefont{Kimura}}\bibnamefont{\emph{
et~al.}},
  \bibinfo{journal}{Phys. Rev. B}
  \textbf{\bibinfo{volume}{68}}, \bibinfo{eid}{060403}
  (\bibinfo{year}{2003}{\natexlab{b}}).

\bibitem[{\citenamefont{Kimura et~al.}(2003{\natexlab{a}})\citenamefont{Kimura,
  Goto, Shintani, Ishizaka, Arima, and Tokura}}]{kimura}
\bibinfo{author}{\bibfnamefont{T.}~\bibnamefont{Kimura}}\bibnamefont{\emph{
et~al.}},
  \bibinfo{journal}{Nature} \textbf{\bibinfo{volume}{426}}, \bibinfo{pages}{55}
  (\bibinfo{year}{2003}{\natexlab{a}}).

\bibitem[{\citenamefont{Lottermoser et~al.}(2004)\citenamefont{Lottermoser,
  Lonkai, Amann, Hohlwein, Ihringer, and Fiebig}}]{lotter}
\bibinfo{author}{\bibfnamefont{Th.}~\bibnamefont{Lottermoser}}\bibnamefont{\emph{
et~al.}},
 \bibinfo{journal}{Nature} \textbf{\bibinfo{volume}{430}},
  \bibinfo{pages}{541} (\bibinfo{year}{2004}).

\bibitem[{\citenamefont{Fiebig}(2005)}]{fiebig:review}
\bibinfo{author}{\bibfnamefont{M.}~\bibnamefont{Fiebig}}, \bibinfo{journal}{J.
  Phys. D}, \textbf{\bibinfo{volume}{38}}, \bibinfo{pages}{R123}
  (\bibinfo{year}{2005}).

\bibitem[{\citenamefont{Goto et~al.}(2004)\citenamefont{Goto et al.}}]{goto}
\bibinfo{author}{\bibfnamefont{T.}~\bibnamefont{Goto}}\bibnamefont{\emph{
et~al.},
  \bibinfo{journal}{Phys. Rev. Lett.}
  \textbf{\bibinfo{volume}{92}}, \bibinfo{eid}{257201}
  (\bibinfo{year}{2004}}).

\bibitem[{\citenamefont{Kimura et~al.}((2005))\citenamefont{Kimura, Lawes,
  Goto, Tokura, and Ramirez}}]{kimura3}
\bibinfo{author}{\bibfnamefont{T.}~\bibnamefont{Kimura}}\bibnamefont{\emph{
et~al.}},
  \bibinfo{journal}{Phys. Rev. B} \textbf{\bibinfo{volume}{71}},
  \bibinfo{pages}{224425} (\bibinfo{year}{2005}).

\bibitem[{\citenamefont{Mostovoy}(2006)}]{mostovoy:067601}
\bibinfo{author}{\bibfnamefont{M.}~\bibnamefont{Mostovoy}},
  \bibinfo{journal}{Phys. Rev. Lett.} \textbf{\bibinfo{volume}{96}},
  \bibinfo{eid}{067601}  (\bibinfo{year}{2006}).


\bibitem[{\citenamefont{Kenzelmann et~al.}(2005)\citenamefont{Kenzelmann,
  Harris, Jonas, Broholm, Schefer, Kim, Zhang, Cheong, Vajk, and Lynn}}]{kenz}
\bibinfo{author}{\bibfnamefont{M.}~\bibnamefont{Kenzelmann}}\bibnamefont{\emph{
et~al.}},
  \bibinfo{journal}{Phys. Rev. Lett.} \textbf{\bibinfo{volume}{95}},
  \bibinfo{eid}{087206} (\bibinfo{year}{2005}).

\bibitem[{\citenamefont{Quezel et~al.}(1977)\citenamefont{Quezel, Tcheou,
  Rossat-Mignod, Quezel, and Roudaut}}]{Quezel}
\bibinfo{author}{\bibfnamefont{S.}~\bibnamefont{Quezel}}\bibnamefont{\emph{
et~al.}},
  \bibinfo{journal}{Physica B\&C} \textbf{\bibinfo{volume}{86-88}},
  \bibinfo{pages}{916} (\bibinfo{year}{1977}).

\bibitem[{\citenamefont{Plumer and Walker}(1982)}]{PW}
\bibinfo{author}{\bibfnamefont{M.~L.}~\bibnamefont{Plumer}} \bibnamefont{and}
  \bibinfo{author}{\bibfnamefont{M.~B.}~\bibnamefont{Walker}},
  \bibinfo{journal}{J. Phys. C}
  \textbf{\bibinfo{volume}{15}}, \bibinfo{pages}{7181} (\bibinfo{year}{1982}).

\bibitem[{\citenamefont{Walker}(1980)}]{walker}
\bibinfo{author}{\bibfnamefont{M.~B.}~\bibnamefont{Walker}},
  \bibinfo{journal}{Phys. Rev. B}
  \textbf{\bibinfo{volume}{22}}, \bibinfo{pages}{1338} (\bibinfo{year}{1980}).

\bibitem[{\citenamefont{Kajimoto et~al.}(2004)\citenamefont{Kajimoto,
  Yoshizawa, Shintani, Kimura, and Tokura}}]{kajimoto}
\bibinfo{author}{\bibfnamefont{R.}~\bibnamefont{Kajimoto}}\bibnamefont{\emph{
et~al.}},
  \space\bibinfo{journal}{Phys. Rev. B}
  \textbf{\bibinfo{volume}{70}}, \bibinfo{eid}{012401}
   (\bibinfo{year}{2004}).

\bibitem[{\citenamefont{Feyerherm et~al.}(2005)\citenamefont{Feyerherm et al.}}]{feyerherm}
\bibinfo{author}{\bibfnamefont{R.}~\bibnamefont{Feyerherm}}\bibnamefont{\emph{
et~al.},
  \bibinfo{journal}{Phys. Rev. B}
  \textbf{\bibinfo{volume}{73}}, \bibinfo{eid}{180401(R)}
  (\bibinfo{year}{2006}}).

\bibitem[{\citenamefont{}()}]{absorption}
\bibinfo{author}{\bibfnamefont{}\bibnamefont{For $^{162}$Dy, $\sigma_{a}$= 194 barn while for natural Dy $\sigma_{a}$= 994 barn, Neutron News $\bf{3}$ 29 (1992).}}







\bibitem[{\citenamefont{Carvajal}(1993)\citenamefont{Carvajal}}]{carvajal}
\bibinfo{author}{\bibfnamefont{J.}~\bibnamefont{Rodriguez-Carvajal}
\bibinfo{journal}{Physica B}
\textbf{\bibinfo{volume}{192}}, \bibinfo{eid}{55}
 (\bibinfo{year}{1993}}).

\bibitem[{\citenamefont{Dudzik et~al.}(2005)\citenamefont{Dudzik et al.}}]{dudzik}
\bibinfo{author}{\bibfnamefont{E.}~\bibnamefont{Dudzik}}~\bibnamefont{\emph{et~al.},
  \bibinfo{journal}{J. Synchrotron Rad.}
  (\bibinfo{year}{2006}}), in press.



\bibitem[{\citenamefont{Brinks}(2005)\citenamefont{Brinks}}]{brinks}
\bibinfo{author}{\bibfnamefont{H.W.}~\bibnamefont{Brinks}}\bibnamefont{\emph{
et~al.},
  \bibinfo{journal}{Phys. Rev. B}
  \textbf{\bibinfo{volume}{63}}, \bibinfo{eid}{094411}
  (\bibinfo{year}{2001}}).

\bibitem[{\citenamefont{Noda et~al.}(2006)}]{noda}
\bibinfo{author}{\bibfnamefont{K.}~\bibnamefont{Noda}}\bibnamefont{\emph{
et~al.}},
  \bibinfo{journal}{cond-matt}/\textbf{\bibinfo{volume}{0512139}}
   (\bibinfo{year}{2005}).

\bibitem[Harris and Lawes, 2006]{harris2}
Harris, A. and Lawes, G. (2006).
\newblock {\em The Handbook of Magnetism and Advanced Magnetic Materials}.
\newblock Wiley.












\end{thebibliography}
\end{document}